\newif\ifreview 
\newif\ifarxiv \newcommand{\arxiv}{\arxivtrue}
\newif\ifcamera 
\newif\ifrebuttal 
\arxiv

\documentclass[10pt,twocolumn,letterpaper]{article}

\usepackage[pagenumbers]{cvpr}

\usepackage{graphicx}
\usepackage{cuted}
\usepackage{listings}
\usepackage{amsmath}
\usepackage{bm}
\usepackage{amssymb}
\usepackage{booktabs}
\usepackage{enumitem}
\usepackage{pifont}
\usepackage{xcolor}
\usepackage{mathtools}
\newcommand{\cmark}{\ding{51}}%
\newcommand{\xmark}{\ding{55}}%

\newcommand{\fidg}{$\text{FID}_g$~}
\newcommand{\fidk}{$\text{FID}_k$~}
\newcommand{\distk}{$\text{Dist}_k$~}
\newcommand{\distg}{$\text{Dist}_g$~}
\newcommand{\websitelink}{https://edge-dance.github.io/}
\newcommand{\website}{\href{\websitelink}{website}}

\usepackage[pagebackref,breaklinks,colorlinks]{hyperref}

\usepackage[capitalize]{cleveref}
\crefname{section}{Sec.}{Secs.}
\Crefname{section}{Section}{Sections}
\Crefname{table}{Table}{Tables}
\crefname{table}{Tab.}{Tabs.}

\begin{document}

\title{EDGE: Editable Dance Generation From Music}

\author{Jonathan Tseng, Rodrigo Castellon, C. Karen Liu\\
Stanford University\\
{\tt\small \{jtseng20,rjcaste,karenliu\}@cs.stanford.edu}
}
\maketitle
\begin{strip}
\centering
    \vspace{-50pt}
    \includegraphics[width=\textwidth]{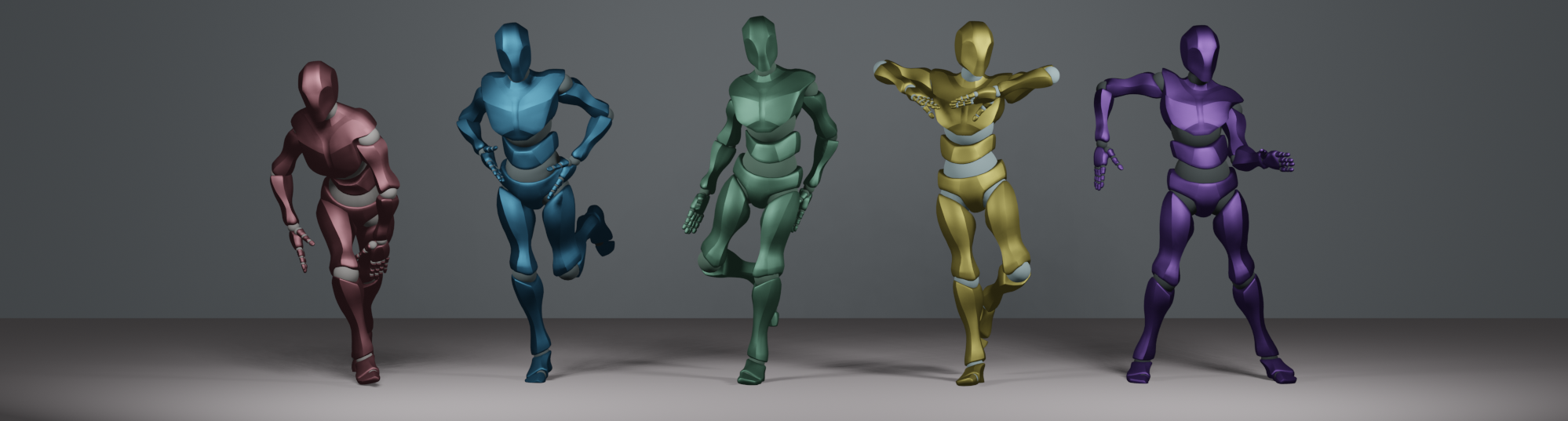}
    \captionof{figure}{EDGE generates diverse, physically plausible dance choreographies conditioned on music.}
    \label{fig:1}
\end{strip}

\begin{abstract}
   Dance is an important human art form, but creating new dances can be difficult and time-consuming.
   In this work, we introduce Editable Dance GEneration (EDGE), a state-of-the-art method for editable dance generation that is capable of creating realistic, physically-plausible dances while remaining faithful to the input music.
   EDGE uses a transformer-based diffusion model paired with Jukebox, a strong music feature extractor, and confers powerful editing capabilities well-suited to dance, including joint-wise conditioning, and in-betweening.
   We introduce a new metric for physical plausibility, and evaluate dance quality generated by our method extensively through (1) multiple quantitative metrics on physical plausibility, beat alignment, and diversity benchmarks, and more importantly, (2) a large-scale user study, demonstrating a significant improvement over previous state-of-the-art methods. Qualitative samples from our model can be found at our \website.
\end{abstract}

\section{Introduction}
\label{sec:intro}

Dance is an important part of many cultures around the world: it is a form of expression, communication, and social interaction \cite{lamothe2019dancing}.
However, creating new dances or dance animations is uniquely difficult because dance movements are expressive and freeform, yet precisely structured by music.
In practice, this requires tedious hand animation or motion capture solutions, which can be expensive and impractical.
On the other hand, using computational methods to generate dances automatically can alleviate the burden of the creation process, leading to many applications: such methods can help animators create new dances or provide interactive characters in video games or virtual reality with realistic and varied movements based on user-provided music.
In addition, dance generation can provide insights into the relationship between music and movement, which is an important area of research in neuroscience \cite{brown2008neuroscience}.

Previous work has made significant progress using machine learning-based methods, but has achieved limited success in generating dances from music that satisfy user constraints.
Furthermore, the evaluation of generated dances is subjective and complex, and existing papers often use quantitative metrics that we show to be flawed.

In this work, we propose Editable Dance GEneration (EDGE), a state-of-the-art method for dance generation that creates realistic, physically-plausible dance motions based on input music.
Our method uses a transformer-based diffusion model paired with Jukebox, a strong music feature extractor.
This unique diffusion-based approach confers powerful editing capabilities well-suited to dance, including joint-wise conditioning and in-betweening.
In addition to the advantages immediately conferred by the modeling choices, we observe flaws with previous metrics and propose a new metric that captures the physical accuracy of ground contact behaviors without explicit physical modeling.
In summary, our contributions are the following:
\vspace{10pt}
\begin{enumerate}[itemsep=0ex,partopsep=1ex,parsep=1ex, topsep=-0ex] 
    \item We introduce a diffusion-based approach for dance generation that combines state-of-the-art performance with powerful \textbf{editing} capabilities and is able to generate \textbf{arbitrarily long} sequences.
    \item We analyze the metrics proposed in previous works and show that they do not accurately represent human-evaluated quality as reported by a large user study.
    \item We propose a new approach to eliminating foot-sliding physical implausibilities in generated motions using a novel Contact Consistency Loss, and introduce Physical Foot Contact Score, a simple new acceleration-based quantitative metric for scoring physical plausibility of generated kinematic motions that requires no explicit physical modeling.
    \item We improve on previous hand-crafted audio feature extraction strategies by leveraging music audio representations from Jukebox~\cite{dhariwal2020jukebox}, a pre-trained generative model for music that has previously demonstrated strong performance on music-specific prediction tasks~\cite{castellon2021codified,donahue2021sheet}.
\end{enumerate}

\vspace{5pt}
This work is best enjoyed when accompanied by our demo samples.
Please see the samples at our \website.

\section{Related Work}
\label{sec:related_work}

\subsection{Human motion generation}

Human motion generation, the problem of automatically generating realistic human motions, is well-studied in computer vision, graphics, and robotics.
Despite its importance and recent progress, it remains a challenging problem, with existing methods often struggling to capture the complexities of physically and stylistically realistic human motion.

Many early approaches fall under the category of \textit{motion matching}, which operates by interpolating between sequences retrieved from a database \cite{holden2020learned}.
While these approaches generate outputs that are physically plausible, their application has been primarily restricted to simple domains such as locomotion.

In recent years, deep neural networks have emerged as a promising alternative method for human motion generation.
These approaches are often capable of generating diverse motions, but often fall short in capturing the physical laws governing human movement or rely on difficult-to-train reinforcement learning solutions \cite{ling2020character, won2022physics}.
Generating human motion conditioned on various inputs---e.g. joystick control~\cite{ling2020character}, class-conditioning~\cite{guo2020action2motion,petrovich2021action}, text-to-motion~\cite{zhang2022motiondiffuse, petrovich2022temos}, seed motions~\cite{rempe2021humor,holden2016deep,duan2022unified,yin2022dance}---is also an active area of study.

\subsection{Dance Generation}

The uniquely challenging task of generating dances stylistically faithful to input music has been tackled by many researchers.
Many early approaches follow a motion retrieval paradigm~\cite{fan2011example,lee2013music,ofli2011learn2dance}, but tend to create unrealistic choreographies that lack the complexity of human dances.
Later works instead synthesize motion from scratch by training on large datasets~\cite{li2021ai,siyao2022bailando,li2022danceformer,kim2022brand,huang2022genre,pu2022music,fan2022bi,au2022choreograph,valle2021transflower} and propose many modeling approaches, including adversarial learning, recurrent neural networks, and transformers.
Despite achieving impressive performance, many such systems are complex~\cite{siyao2022bailando,au2022choreograph,li2022danceformer,chen2021choreomaster}, often involving many layers of conditioning and sub-networks.
In contrast, our proposed method contains a single model trained with a simple objective, yet offers both strong generative and editing capabilities without significant hand-crafted design.
\begin{figure*}
    \centering
    \includegraphics[width=\linewidth]{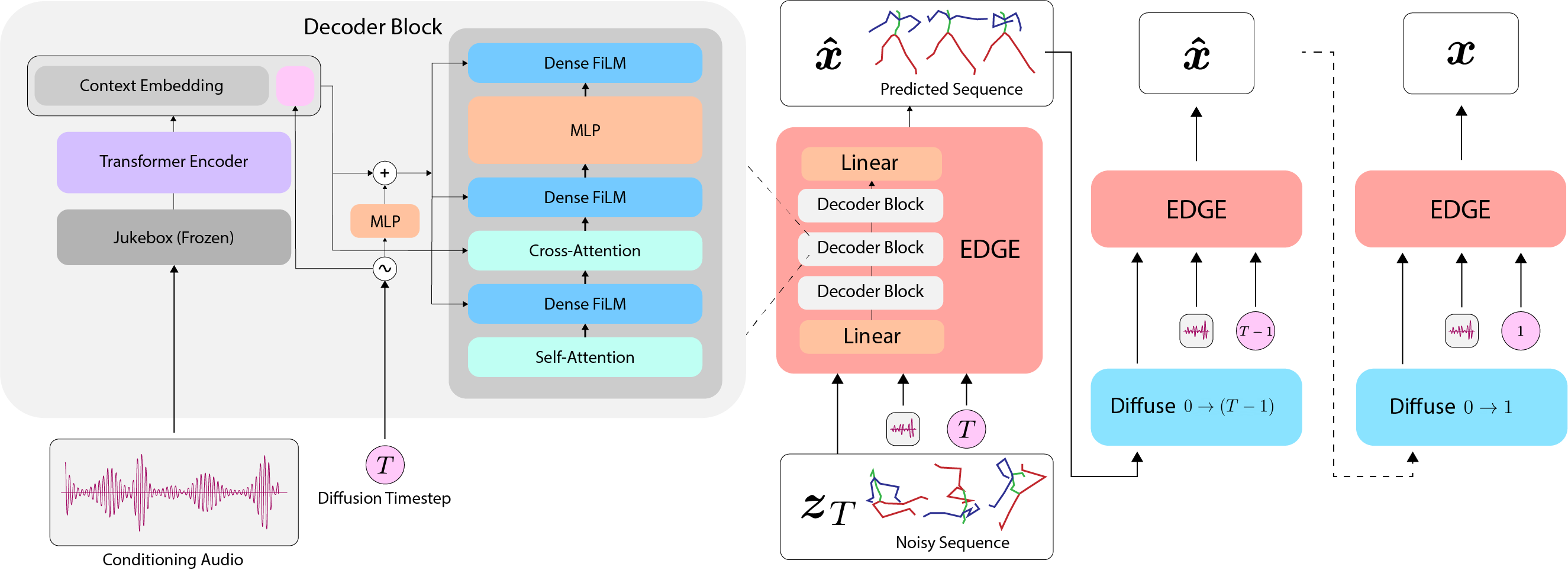}
    \caption{\textbf{EDGE Pipeline Overview:} EDGE learns to denoise dance sequences from time $t=T$ to $t=0$, conditioned on music. Music embedding information is provided by a frozen Jukebox model~\cite{dhariwal2020jukebox} and acts as cross-attention context. EDGE takes a noisy sequence $\bm{z}_T \sim \mathcal{N}(0, \bm{I})$ and produces the estimated final sequence $\bm{\hat x}$, noising it back to $\bm{\hat{z}}_{T-1}$ and repeating until $t=0$.}
    \label{fig:2}
\end{figure*}

\subsection{Generative Diffusion Models}
Diffusion models~\cite{sohl2015deep, ho2020denoising} are a class of deep generative models which learn a data distribution by reversing a scheduled noising process.
In the past few years, diffusion models have been shown to be a promising avenue for generative modeling, exceeding the state-of-the-art in image and video generation tasks~\cite{saharia2022photorealistic, ho2022imagen}.
Much like previous generative approaches like VAE~\cite{kingma2019introduction} and GAN~\cite{goodfellow2020generative}, diffusion models are also capable of conditional generation.
Dhariwal et al.~\cite{dhariwal2021diffusion} introduced classifier guidance for image generation, where the output of a diffusion model may be ``steered" towards a target, such as a class label, using the gradients of a differentiable auxiliary model.
Saharia et al.~\cite{saharia2022palette} proposed to use direct concatenation of conditions for Pix2Pix-like tasks, akin to Conditional GAN and CVAE\cite{mirza2014conditional, sohn2015learning}, and Ho et al.~\cite{ho2022classifier} demonstrated that classifier-free guidance can achieve state-of-the-art results while allowing more explicit control over the diversity-fidelity tradeoff.

Most recently, diffusion-based methods have demonstrated strong performance in generating motions conditioned on text~\cite{tevet2022human,zhang2022motiondiffuse,kim2022flame}.
While the tasks of text-to-motion and music-conditioned dance generation share high-level similarities, the dance generation task suffers more challenging computational scaling (see \cref{par:longform}) and, due to its specialized nature, much lower data availability.

\section{Method} \label{sec:method}

\paragraph{Pose Representation}
 We represent dances as sequences of poses in the 24-joint SMPL format \cite{loper2015smpl}, using the 6-DOF rotation representation \cite{zhou2019continuity} for every joint and a single root translation: $\bm{w} \in \mathbb{R}^{24 \cdot 6 + 3 = 147}$. For the heel and toe of each foot, we also include a binary contact label: $\bm{b} \in \{0,1\}^{2 \cdot 2 = 4}$. The total pose representation is therefore $\bm{x} = \{\bm{b} , \bm{w}\} \in \mathbb{R}^{4+147=151}.$
EDGE uses a diffusion-based framework to learn to synthesize sequences of $N$ frames, $\bm{x} \in \mathbb{R}^{N\times 151}$, given arbitrary music conditioning $\bm{c}$.

\paragraph{Diffusion Framework}
We follow the DDPM~\cite{ho2020denoising} definition of diffusion as a Markov noising process with latents $\{\bm{z}_t\}_{t=0}^T$ that follow a forward noising process $q(\bm{z}_t | \bm{x})$, where $\bm{x} \sim p(\bm{x})$ is drawn from the data distribution.
The forward noising process is defined as
\begin{equation} \label{eq:1}
{
q(\bm{z}_{t} | \bm{x}) \sim \mathcal{N}(\sqrt{\bar{\alpha}_{t}}\bm{x}, (1 - \bar{\alpha}_{t})\bm{I}),
}
\end{equation}
where $\bar{\alpha}_{t} \in (0,1)$ are constants which follow a monotonically decreasing schedule such that when $\bar{\alpha}_{t}$ approaches 0, we can approximate $\bm{z}_{T} \sim \mathcal{N}(0,\bm{I})$.

In our setting with paired music conditioning $\bm{c}$, we reverse the forward diffusion process by learning to estimate $\bm{\hat{x}}_\theta(\bm{z}_t, t, \bm{c}) \approx \bm{x}$ with model parameters $\theta$ for all $t$.
We optimize $\theta$ with the ``simple'' objective introduced in Ho et al.~\cite{ho2020denoising}:
\begin{equation} \label{eq:2}
{
\mathcal{L}_\text{simple} = \mathbb{E}_{\bm{x}, t}\left[\| \bm{x} - \bm{\hat{x}}_\theta(\bm{z}_t, t, \bm{c})\|_2^2\right].
}
\end{equation}
From now on, we refer to $\bm{\hat{x}}_\theta(\bm{z}_t, t, \bm{c})$ as $\bm{\hat{x}}(\bm{z}_t,\bm{c})$ for ease of notation.

\paragraph{Auxiliary losses} 
Auxiliary losses are commonly used in kinematic motion generation settings to improve physical realism in the absence of true simulation \cite{tang2022real, shi2020motionet, petrovich2021action}.
In addition to the reconstruction loss $\mathcal{L}_\text{simple}$, we adopt auxiliary losses similar to those in Tevet et al.~\cite{tevet2022human}, which encourage matching three aspects of physical realism: joint positions (\cref{eq:3}), velocities (\cref{eq:4}), and foot velocities via our Contact Consistency Loss (\cref{eq:5}).
\begin{equation} \label{eq:3}
{
\mathcal{L}_\text{joint} = \frac{1}{N} \sum_{i =1}^{N} \| FK(\bm{x}^{(i)}) - FK(\bm{\hat{x}}^{(i)})\|_{2}^{2} %
} 
\end{equation}
\begin{equation} \label{eq:4}
{
\mathcal{L}_\text{vel} = \frac{1}{N-1} \sum_{i =1}^{N-1} \| (\bm{x}^{(i+1)} - \bm{x}^{(i)}) - (\bm{\hat{x}}^{(i+1)} - \bm{\hat{x}}^{(i)})\|_{2}^{2}
}
\end{equation}

\begin{equation} \label{eq:5}
{
\mathcal{L}_\text{contact} = \frac{1}{N-1} \sum_{i =1}^{N-1} \| (FK(\bm{\hat{x}}^{(i+1)}) - FK(\bm{\hat{x}}^{(i)})) \cdot \bm{\hat{b}}^{(i)}\|_{2}^{2}
},
\end{equation}
where $FK(\cdot)$ denotes the forward kinematic function which converts joint angles into joint positions (though it only applies to the foot joints in \cref{eq:5}) and the $(i)$ superscript denotes the frame index.
In the Contact Consistency Loss, $\bm{\hat{b}}^{(i)}$ is the model's own prediction of the binary foot contact label's portion of the pose at each frame $i$.
While this formulation is similar to the foot skate loss terms in previous works~\cite{tevet2022human, tang2022real}, where foot velocity is penalized in frames where the ground truth motion exhibits a static foot contact, our Contact Consistency Loss formulation differs in that it encourages the model to (1) predict foot contact, and (2) \textit{maintain consistency with its own predictions}.
We find that this formulation significantly improves the realism of generated motions (see \cref{par:exp_physical_plausibility}).

Our overall training loss is the weighted sum of the simple objective and the auxiliary losses:
\begin{equation} \label{eq:6}
{
\mathcal{L} = \mathcal{L}_\text{simple} + \lambda_\text{pos}\mathcal{L}_\text{pos} + \lambda_\text{vel}\mathcal{L}_\text{vel} + \lambda_\text{contact}\mathcal{L}_\text{contact}.
}
\end{equation}

\paragraph{Sampling and Guidance}
At each of the denoising timesteps $t$, EDGE predicts the denoised sample and noises it back to timestep $t-1$ : $\bm{\hat{z}}_{t-1} \sim q(\bm{\hat{x}}_\theta(\bm{\hat{z}}_t, \bm{c}), t-1)$, terminating when it reaches $t=0$ (\cref{fig:2}, right).
We train our model using classifier-free guidance \cite{ho2022classifier}, which is commonly used in diffusion-based models \cite{saharia2022photorealistic, rombach2022high, tevet2022human, kim2022flame, zhang2022motiondiffuse}.
Following Ho et al.~\cite{ho2022classifier}, we implement classifier-free guidance by randomly replacing the conditioning with $\bm{c} = \emptyset$ during training with low probability (e.g. 25\%).
Guided inference is then expressed as the weighted sum of unconditionally and conditionally generated samples:
\begin{equation} \label{eq:7}
{
\bm{\tilde{x}}(\bm{\hat{z}}_t, \bm{c}) = w \cdot \bm{\hat{x}}(\bm{\hat{z}}_t, \bm{c}) + (1 - w) \cdot \bm{\hat{x}}(\bm{\hat{z}}_t, \emptyset).
}
\end{equation}
At sampling time, we can amplify the conditioning $\bm{c}$ by choosing a guidance weight $w > 1$.

\begin{figure}
    \centering
    \includegraphics[width=\linewidth]{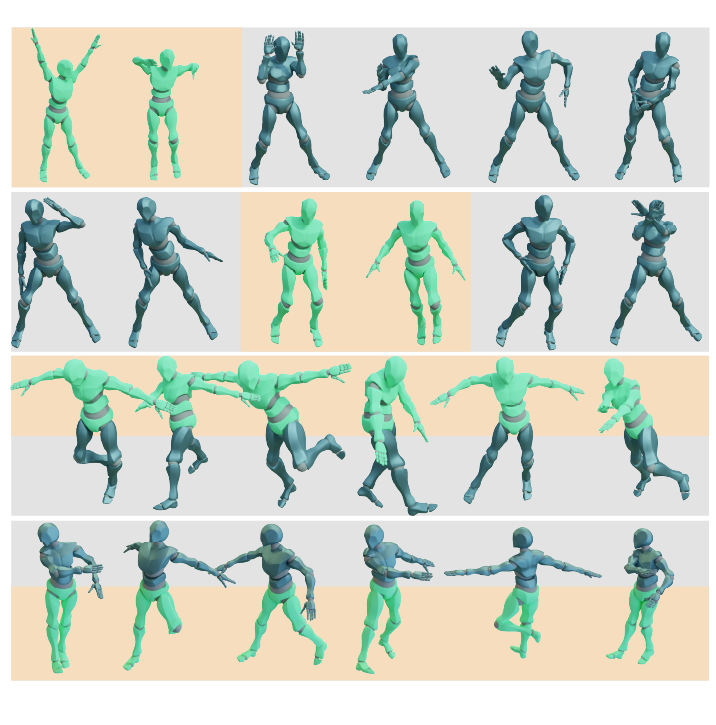}
    \caption{EDGE allows the user to specify both temporal and joint-wise constraints. Constraint joints / frames are highlighted in green and tan, generated joints / frames are in blue and gray.  Pictured, top to bottom: dance completion from seed motion, dance that hits a specified keyframe mid-choreography, completion from specified upper-body joint angles, completion from specified lower-body joint angles and root trajectory.}
    \label{fig:3}
\end{figure}

\paragraph{Editing}
To enable editing for dances generated by EDGE, we use the standard masked denoising technique from diffusion image inpainting~\cite{lugmayr2022repaint}, and more recently text-to-motion models~\cite{kim2022flame, tevet2022human}. EDGE supports any combination of temporal and joint-wise constraints, shown in \cref{fig:3}. Given a joint-wise and/or temporal constraint $\bm{x}^{\text{known}}$ with positions indicated by a binary mask $\bm{m}$, we perform the following at every denoising timestep:
\begin{equation} \label{eq:8}
    \bm{\hat{z}}_{t-1} \coloneqq \bm{m} \odot q(\bm{x}^{\text{known}}, t-1) + (1 - \bm{m}) \odot \bm{\hat{z}}_{t-1},
\end{equation}
where $\odot$ is the Hadamard (element-wise) product, replacing the known regions with forward-diffused samples of the constraint. This technique allows editability at inference time with no special training processes necessary.

For example, a user can perform motion in-betweening by providing a reference motion $\bm{x}^{\text{known}} \in \mathbb{R}^{N \times 151}$ and a mask $\bm{m} \in \{0,1\}^{N \times 151}$, where $\bm{m}$ is all 1's in the first and last $n$ frames and 0 everywhere else.
This would result in a sequence $N$ frames long, where the first and last $n$ frames are provided by the reference motion and the rest is filled in with a plausible ``in-between" dance that smoothly connects the constraint frames, for arbitrary $2n < N$.
This editing framework provides a powerful tool for downstream applications, enabling the generation of dances that precisely conform to arbitrary constraints.

\begin{figure}
    \centering
    \includegraphics[width=0.8\linewidth]{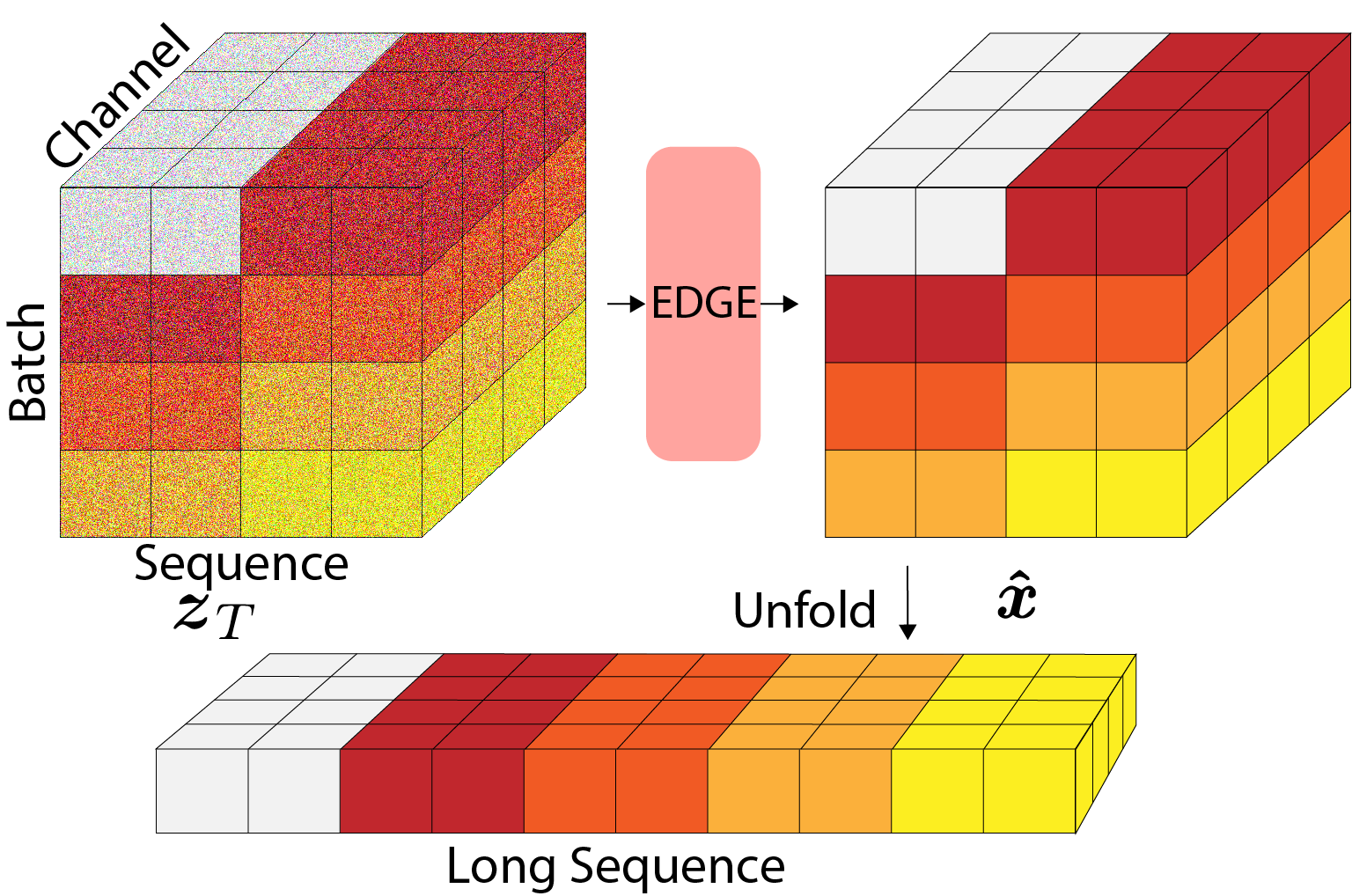}
    \caption{Although EDGE is trained on 5-second clips, it can generate choreographies of any length by imposing temporal constraints on batches of sequences.
    In this example, EDGE constrains the first 2.5 seconds of each sequence to match the last 2.5 seconds of the previous one to generate a 12.5-second clip, as represented by the temporal regions of distinct clips in the batch that share the same color.}
    \label{fig:4}
    \vspace{-15pt}
\end{figure}

\paragraph{Long-form sampling}
\label{par:longform}
The ability to synthesize sequences of arbitrary length, often many minutes long, is critical to the task of dance generation.
However, since EDGE generates every frame of a dance sequence at once, naively increasing the maximum sequence length incurs a linear increase in computational cost.
Moreover, dance generation requires that the conditioning $\bm{c}$ match the motion sequence in length, causing further scaling of memory demands, which is especially severe in the case of embeddings from large models like Jukebox~\cite{dhariwal2020jukebox}.
To approach the challenge of long-form generation, EDGE leverages its editability to enforce temporal consistency between multiple sequences such that they can be concatenated into a single longer sequence.
Refer to \cref{fig:4} for a depiction of this process.

\paragraph{Model}
Our model architecture is illustrated in \cref{fig:2}.
We adopt a transformer decoder architecture, which processes music conditioning projected to the transformer dimension with a cross-attention mechanism following Saharia et al.~\cite{saharia2022photorealistic}.
Timestep information is incorporated both with a token concatenated with the music conditioning and feature-wise linear modulation (FiLM) \cite{perez2018film}.

\paragraph{Music Audio Features}

Past work~\cite{li2021ai,siyao2022bailando,li2022danceformer,kim2022brand,huang2022genre,pu2022music,fan2022bi,au2022choreograph,valle2021transflower} has largely focused on advancing the generative modeling approach for the dance generation problem, with little focus on the representation of the music signal itself, which we argue is equally important.
Indeed, state-of-the-art results in the text-to-image domain have found that scaling the \textit{text encoder} is more important for performance than scaling the diffusion model \cite{saharia2022photorealistic}.

In this vein, recent work~\cite{castellon2021codified,donahue2021sheet} in music information retrieval has demonstrated that Jukebox~\cite{dhariwal2020jukebox}, a large GPT-style model trained on 1M songs to generate raw music audio, contains representations that serve as strong features for music audio prediction tasks.
We take inspiration from these advances and extract Jukebox features as the conditioning input for our diffusion model, and develop a new memory-efficient implementation that enables near real-time extraction on a single commodity GPU (see \cref{sec:appendix_juke_mem} for details).

\section{Experiments}

\paragraph{Dataset}
In this work, we use AIST++~\cite{li2021ai}, a dataset consisting of 1,408 high-quality dance motions paired to music from a diverse set of genres.
We re-use the train/test splits provided by the original dataset. All training examples are cut to 5 seconds, 30 FPS.
\vspace{-10pt}

\paragraph{Baselines}

Among recent state-of-the-art dance generation methods \cite{li2021ai,siyao2022bailando,li2022danceformer,kim2022brand,huang2022genre,pu2022music,fan2022bi,au2022choreograph,valle2021transflower}, we select the following baselines:
\begin{itemize}
    \item FACT \cite{li2021ai}, an autoregressive model introduced together with the AIST++ dataset
    \item \textit{Bailando} \cite{siyao2022bailando}, a follow-up approach that achieves the strongest qualitative performance to date.
\end{itemize}

\vspace{-10pt}

\paragraph{Implementation details}

Our final model has 49M total parameters, and was trained on 4 NVIDIA A100 GPUs for 16 hours with a batch size of 512.

For long-form generation, we use 5-second slices and choose to enforce consistency for overlapping 2.5-second slices by interpolating between the two slices with linearly decaying weight.
We find that this simple approach is sufficient to result in smooth, consistent generation, as demonstrated throughout our \website.

We evaluate two separate feature extraction strategies. The ``baseline'' strategy uses the popular audio package \textit{librosa}~\cite{mcfee2015librosa} to extract beats and accompanying audio features using the same code as in Li et al.~\cite{li2021ai} (when the ground truth BPM is not known, we estimate it with \textit{librosa}). The ``Jukebox''~\cite{dhariwal2020jukebox} extraction strategy follows Castellon et al.~\cite{castellon2021codified}, extracting representations from Jukebox and downsampling to 30 FPS to match our frame rates for motion data.

\subsection{Comparison to Existing Methods}

In this section, we compare our proposed model to several past works on (1) human evaluations, (2) our proposed physical plausibility metric, (3) beat alignment scores, (4) diversity, and (5) performance on in-the-wild music.

\begin{table*}[t]
  \centering
  \begin{tabular}{lcccccccc}
    \toprule
    Method & Elo $\uparrow$ & EDGE Win Rate & PFC $\downarrow$ & Beat Align. $\uparrow$ & \distk $\xrightarrow[]{}$ & \distg $\xrightarrow[]{}$ & Fixed Bones & Editing \\
    \midrule
    EDGE ($w=2$) & \textbf{1751} & N/A & \textbf{1.5363} & 0.26 & 9.48 & 5.72 & \cmark & \cmark \\
    EDGE ($w=1$) & 1601 & 58.0\% $\pm$ 3.8\% & 1.6545 & \textbf{0.27} & \textbf{10.58} & \textbf{7.62} & \cmark & \cmark \\
    \midrule
    \textit{Bailando} & 1397 & 91.1\% $\pm$ 5.9\% & 1.754 & 0.23 & 7.92 & 7.72 & \xmark & \xmark \\
    FACT & 1325 & 90.0\% $\pm$ 7.0\% & 2.2543 & 0.22 & 10.85 & 6.14 & \cmark & \xmark \\
    \midrule Ground Truth & 1653 & 65.7\% $\pm$ 11.1\% & 1.332 & 0.24 & 10.61 & 7.48 & \cmark & N/A \\
    \bottomrule
\end{tabular}
    \caption{We compare our method against FACT~\cite{li2021ai} and \textit{Bailando}~\cite{siyao2022bailando}.
    In the table, $w$ refers to the guidance weight at inference time.
    We evaluate all methods qualitatively via human raters to obtain Elo~\cite{elo1978rating} and Win Rate, and compute the rest of the metrics automatically.
    For reference, the Elo rating system is designed such that a 400 point gap corresponds to a $\sim\hspace{-2pt}90$\% head-to-head win rate, which is reflected in our empirical results (the ``EDGE Win Rate'' column).
    Refer to \href{https://en.wikipedia.org/wiki/Elo_rating_system}{the Wikipedia page} for the mathematical details.
    Error bars for ``EDGE Win Rate'' correspond to a 95\% confidence interval (see \cref{sec:appendix_user_study} for details).
    $\uparrow$ means higher is better, $\downarrow$ means lower is better, and $\xrightarrow[]{}$ means closer to ground truth is better.
    }
    \label{tab:main_table}
  \end{table*}
\vspace{-10pt}
\paragraph{Human Evaluations}
\label{par:human_evaluations}

To obtain the results in \cref{tab:main_table}, we recruited 147 human raters via Prolific, a crowdsourcing platform for recruiting study participants.
In aggregate, raters evaluated 11,610 pairs of clips randomly sampled from our models, ground truth, baseline models, or a checkpoint from our model training (as seen in \cref{par:fidk}).
All dances were rendered in the same setting with music sampled uniformly at random.
Raters were asked to select the dance that ``looked and felt better overall''.
In addition to computing the raw win rate of our method when evaluated directly against competing methods shown in the table, we also computed each method's Elo rating, which, unlike flat win rate, is able to simultaneously capture the relative quality of more than 2 models~\cite{elo1978rating}.

The study reveals that human raters overwhelmingly prefer the dances generated by EDGE over previous methods, and  even favor EDGE dances over real dances.

For more information about the exact details of our user studies, see \cref{sec:appendix_user_study}.

\paragraph{Physical Plausibility}
\label{par:exp_physical_plausibility}

Any generated dance should be physically plausible; otherwise, its downstream applicability is dramatically limited.
Previous works evaluate plausibility of foot-ground contact by measuring foot sliding ~\cite{taheri2022goal, he2022nemf}; however, dance is unique in that sliding is not only common but integral to many choreographies.
This reflects a need for a metric that can measure the realism of foot-ground contact that does not assume that feet should exhibit static contact at all times.
In this work, we propose a new metric to evaluate physical plausibility, Physical Foot Contact score (PFC), that we believe captures this concept well.

PFC is a physically-inspired metric that requires no explicit physical modeling.
Our metric arises from two simple, related observations:
\begin{enumerate}[itemsep=0ex,partopsep=0ex,parsep=0ex]
    \item On the horizontal (xy) plane, any center of mass (COM) acceleration must be due to static contact between the feet and the ground.
    Therefore, either at least one foot is stationary on the ground or the COM is not accelerating.
    \item On the vertical (z) axis, any \textit{positive} COM acceleration must be due to static foot contact.
\end{enumerate}

Therefore, we can represent adherence to these conditions as an average over time of the below expression, scaled to normalize acceleration:

\begin{equation} \label{eq:9}
{
s^i = ||\overline{\bm{a}}^i_{\text{COM}}|| \cdot ||\mathbf{v}^i_{\text{Left Foot}}|| \cdot ||\mathbf{v}^i_{\text{Right Foot}}||
},
\end{equation}
\begin{equation} \label{eq:10}
{
PFC = \frac{1}{N \cdot \max\limits_{1\leq j \leq N} ||\bm{\overline{a}}^{j}_{\text{COM}}||} \sum_{i=1}^{N} s^i
},
\end{equation}
where  $$\overline{\bm{a}}_{\text{COM}}^i = \begin{pmatrix} a_{\text{COM},x}^i \\ a_{\text{COM},y}^i \\ \max(a_{\text{COM},z}^i, 0) \end{pmatrix}$$

and the $i$ superscript denotes the frame index.

\begin{table}
  \centering
  \begin{tabular}{lcc}
    \toprule
    Method (In-the-Wild) & Elo $\uparrow$ & EDGE Win Rate \\
    \midrule EDGE & 1747 & 53.8\% $\pm$ 11.1\% \\
    \hspace{4pt} w/o Jukebox & 1603 & 83.3\% $\pm$ 8.3\% \\
    \textit{Bailando} & 1222 & 82.4\% $\pm$ 5.5\% \\
    FACT & 1367 & 89.3\% $\pm$ 4.4\% \\
    \bottomrule
    \end{tabular}
    \caption{We test our model on in-the-wild music and ablate Jukebox features in this setting.}
    \label{tab:ood_table}
\end{table}

In our results (\cref{tab:main_table}), we find that our method attains a greater level of physical plausibility than previous methods, and approaches the plausibility of ground truth motion capture data.

Maintaining fixed bone length is another important aspect of physical plausibility.
Our method operates in the reduced coordinate (joint angle) space, which guarantees fixed bone length. However, methods that operate in the joint Cartesian space, such as \emph{Bailando}, can produce significantly varying bone lengths.
For example, on average, bone lengths change up to $\pm20\%$ over the course of dance sequences generated by \textit{Bailando}.

\vspace{-10pt}
\paragraph{Beat Alignment Scores}

Our experiments evaluate the tendency of our generated dances to follow the beat of the music, following previous work~\cite{siyao2022bailando} for the precise implementation of this metric.
The results demonstrate that EDGE outperforms past work, including \textit{Bailando}, which includes a reinforcement learning module that explicitly optimizes beat alignment.
We further examine the robustness of this metric in \cref{par:beat_alignment_discussion}.
\vspace{-10pt}
\paragraph{Diversity}

Diversity metrics are computed following the methods of previous work~\cite{li2021ai,siyao2022bailando}, which measure the distributional spread of generated dances in the ``kinetic'' (Dist$_k$) and ``geometric'' (Dist$_g$) feature spaces, as implemented by \textit{fairmotion} \cite{gopinath2020fairmotion, muller2005efficient, onuma2008fmdistance}.
We compute these metrics on 5-second dance clips produced by each approach.
Given that the ultimate goal of dance generation is to automatically produce dances that emulate the ground truth distribution, we argue that models should aim to \emph{match} the scores of the ground truth distribution rather than \emph{maximize} their absolute values.
Indeed, past work has found that jittery dances result in high diversity scores, in some cases exceeding the ground truth \cite{li2020learning, li2021ai}.

At low guidance weight ($w=1$), EDGE achieves a level of diversity that closely matches that of the ground truth distribution while attaining state-of-the-art qualitative performance.
At high guidance weight ($w=2$), EDGE produces dances of markedly higher fidelity, reflected by a significantly higher Elo rating, while trading off diversity.
These results reflect those of past studies, which show that sampling using guidance weights $w > 1$ significantly increases fidelity at the cost of diversity \cite{ho2022classifier, saharia2022photorealistic}.
This simple control over the diversity-fidelity tradeoff via the modulation of a single scalar parameter provides a powerful tool for downstream applications.

\paragraph{In-the-Wild Music}

While past approaches achieve strong results on AIST++, these results are not necessarily indicative of the models' ability to generalize to in-the-wild music inputs.
In order to evaluate generalization, we tested our proposed method and the baseline approaches on a diverse selection of popular songs from YouTube.

\begin{table}
  \centering
  \begin{tabular}{lcc}
    \toprule
    Method & EDGE Win Rate & PFC $\downarrow$ \\
    \midrule
    EDGE & N/A & 1.5363 \\
    \hspace{4pt} w/o CCL & 61.8\% $\pm$ 7.3\% & 3.0806 \\
    Ground Truth & 40.6\% $\pm$ 7.4\% & 1.332 \\
    \bottomrule
    \end{tabular}
    \caption{We ablate our contact consistency loss (CCL) and study its impact on qualitative and quantitative physical plausibility.
    While EDGE wins more on average against ground truth on overall quality evaluation (\cref{tab:main_table}), the user study for this table specifically asks about physical plausibility, and we find that ground truth still performs favorably compared to EDGE.}
    \label{tab:physical_plausibility_ablation}
\end{table}

The results, as shown in \cref{tab:ood_table}, demonstrate that our proposed method continues to perform well on in-the-wild music.
We find through ablation that Jukebox features are critical for performance in this setting, bringing human-rated quality almost to par with in-distribution music; furthermore, we find that our approach continues to beat baselines in human evaluations for dance quality.
We note that \emph{Bailando} has an additional fine-tuning procedure available to improve performance on in-the-wild samples by training its actor-critic component.
While we evaluated \emph{Bailando} directly without additional fine-tuning, we found that human raters significantly preferred in-the-wild EDGE samples over in-distribution \emph{Bailando} samples.
We interpret these results as evidence that our proposed method is able to successfully generalize to in-the-wild music inputs.

\subsection{Additional Evaluation}

\paragraph{Editing}

We find that our model is capable of in-betweening, motion continuation, joint-conditioned generation, and long-form generation with quality on par with unconstrained samples.
Please see our \website ~for demo examples.

\vspace{-10pt}
\paragraph{Physical Plausibility}

We test the soundness of our PFC metric and ablate our Contact Consistency Loss (CCL) term (see \cref{eq:5}) using a human evaluation study.
Raters were shown pairs of dances from the ground truth distribution, EDGE with CCL, and EDGE without CCL, and asked which dance looked more physically plausible.
The results (\cref{tab:physical_plausibility_ablation}) show that CCL noticeably improves both the PFC metric and qualitative physical plausibility user evaluations, winning 61.8\% of matchups against the version without CCL, and coming close to parity with ground truth samples.
The results also indicate that PFC tracks well with human perception of physical plausibility.

For a discussion of limitations and implementation details of PFC, please see \cref{par:physical_plausibility} and \cref{sec:appendix_pfc}, respectively.

\subsection{FID Results}
A reasonable solution to the problem of \textit{overall evaluation} is to compute the difference between an empirical distribution of generated dances and a ground truth distribution.
Several past works follow this intuition and automatically evaluate dance quality with Frechet Distance metrics~\cite{li2021ai,li2020learning,siyao2022bailando}. As part of our experiments, we conduct a two-pronged analysis into the prevailing metrics, ``FID$_k$" and ``FID$_g$"~\cite{li2021ai}, which compute the difference between distributions of heuristically extracted motion features, and show them to be unsound for the AIST++ dataset.
\vspace{-10pt}

\paragraph{``FID$_g$"}

\begin{figure}
    \centering
    \includegraphics[width=\linewidth]{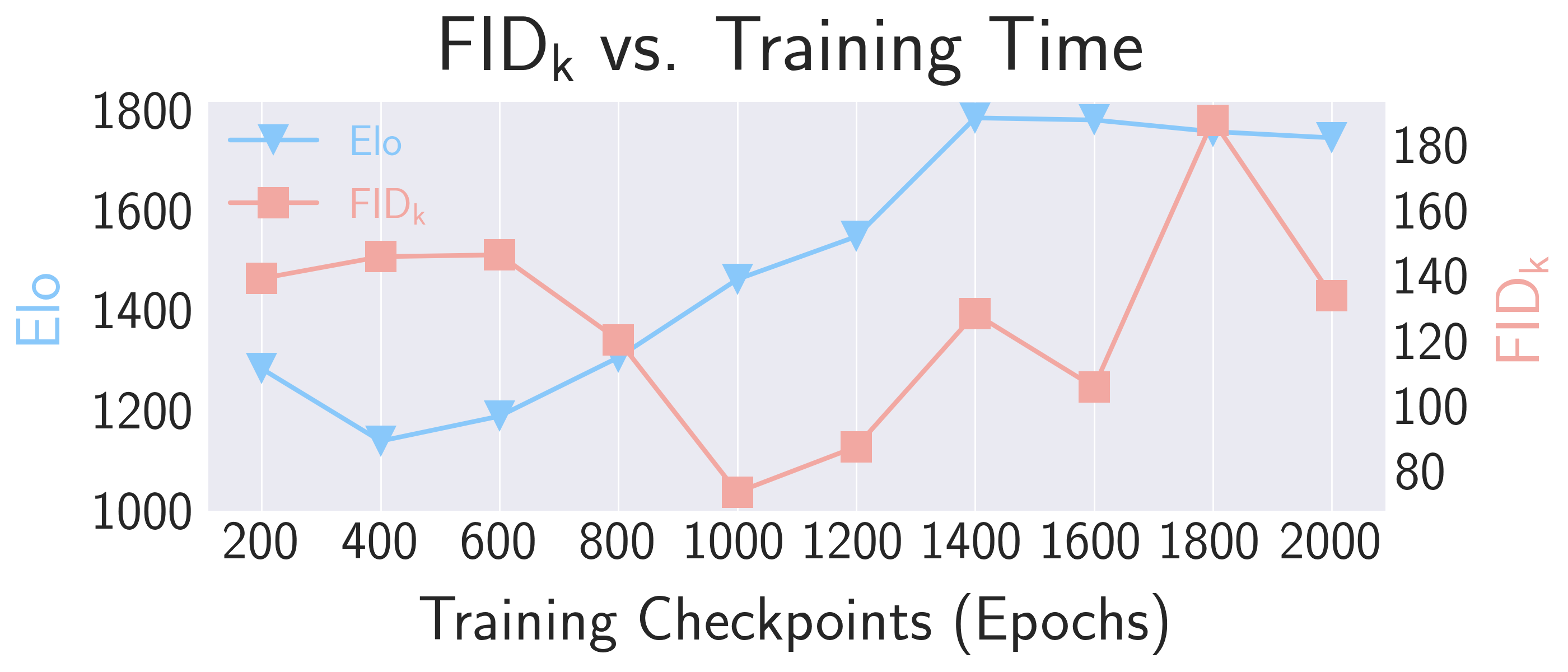}
    \caption{We plot \fidk over the course of model training and find that it is inconsistent with overall quality evaluations.}
    \label{fig:5}
    \vspace{-10pt}
\end{figure}

We compute \fidg on the ground truth AIST++ dataset following the implementation in Li et al.~\cite{li2021ai}.
Specifically, we compute metrics on the test set against the training set and obtain 
a score of 41.4, a stark increase from metrics obtained from generated outputs: running on FACT outputs gives 12.75, running on \textit{Bailando} outputs gives 24.82, and running on our final model's outputs gives 23.08.
Given that FACT performs the best in ``FID$_g$" but performs the worst according to user studies and that all three models perform significantly better than ground truth, we conclude that the ``FID$_g$" values are unreliable as a measure of quality on this dataset.
\vspace{-10pt}
\paragraph{``FID$_k$"}
\label{par:fidk}
We take 10 checkpoints throughout the training process for our model and poll raters on the overall quality of dances sampled from each checkpoint in a round-robin tournament.
Intuitively, this should result in a consistent-to-monotonic improvement in both qualitative performance and in the quantitative quality metrics.
However, we observe that this is not the case.
In \cref{fig:5}, we show the results of this experiment with our proposed model's FID$_k$.

While the qualitative performance of the generated motion improves considerably as we train the model, the \fidk does not improve, and actually significantly worsens during the latter half of training.

\section{Discussion}

Developing automated metrics for the quality of generated dances is a fundamentally challenging undertaking, since the practice of dance is complex, subjective, and even culturally specific.
In this section, we will discuss flaws in automated metrics currently in use and limitations with our proposed PFC metric.

\paragraph{Overall Quality Evaluation}

Our results suggest that despite the current standardized use of FIDs for evaluating dance generation models on the AIST++ dataset, these FID metrics, as they currently stand, are unreliable for measuring quality.
One potential explanation may be that the AIST++ test set does not thoroughly cover the train distribution given its very small size.
Additionally, due to data scarcity, both \fidk and \fidg depend on heuristic feature extractors that only compute superficial features of the data.
By contrast, FID-based metrics in data-rich domains such as image generation (where FID originated) enjoy the use of learned feature extractors (e.g. pre-trained ImageNet models), which are known to directly extract semantically meaningful features \cite{sharif2014cnn}.
Indeed, in the text-to-motion domain, where significantly more paired data is available \cite{guo2022generating}, the standard feature extractors for FID-style metrics depend on deep contrastively learned models \cite{tevet2022human, zhang2022motiondiffuse}.

Our experiments suggest that the current FID metrics can be further improved for evaluating motion sequences.
We believe that the \emph{idea} of evaluating the difference between two featurized distributions of dance motions is not necessarily fundamentally flawed, but that more representative features could result in reliable automatic quality evaluations.

\paragraph{Beat Alignment Scoring}
\label{par:beat_alignment_discussion}

In automated metrics, we should also aim to capture one of the core aspects of a dance performance: its ability to ``follow the beat'' of the song.
Previous work~\cite{li2021ai,siyao2022bailando} has proposed using a beat alignment metric that rewards kinematic beats (local minima of joint speed) that line up with music beats.
In Li et al.~\cite{li2021ai}, the distance is computed per kinematic beat, and in Siyao et al.~\cite{siyao2022bailando} the distance is computed per music beat.

Here, we argue that this metric has some fundamental issues and support this with empirical observations.
Firstly, dance is not strictly about matching local minima in joint speed to beats.
Instead, the beats of the music are a loose guide for timing, rhythm, and transitions between movements and dance steps.
Secondly, while encouraging an overall alignment with the beat of the music is intuitive, the metric as presented in Li et al.~\cite{li2021ai} implies that high-quality dances that contain motions with local minima between beats should be penalized, and the metric as presented in Siyao et al.~\cite{siyao2022bailando} implies that dances that skip over beats should receive corresponding penalties. While each of these metrics holds some truth, neither is strictly correct: a perfectly valid dance which aligns with the music in double time or half time will be scored very poorly by at least one of the two metrics.

Empirically, we observe that in addition to exceeding scores of competing baselines, our beat alignment scores exceed those of ground truth samples.
This potentially suggests that while the metric has served to drive progress on this problem in the past, it is possible that the metric loses its meaning when candidate examples reach quality on par with ground truth.

\paragraph{Physical Plausibility} \label{par:physical_plausibility}

In this work, we introduce PFC, \cref{eq:10}, a physically-inspired metric that specifically targets the challenging issue of foot sliding.
While PFC is intuitive, it is not without its limitations.
In its current form, PFC assumes that the feet are the only joints that experience static contact.
This means that PFC cannot be applied without modification to motions such as gymnastics routines, where a variable number of non-foot (e.g. hand) contacts are integral to their execution.
We believe that a careful analysis of other contact points (e.g. hands in gymnastics) could provide an extension of PFC that is more widely applicable.
Another issue is the assumption that the COM can be accelerated only by static contact: though rare, it is possible to accelerate the COM using friction during extended sliding (which is not present in AIST++).
For more analysis and discussion of PFC, see \cref{sec:appendix_pfc}.

\section{Conclusion and Future Work}

In this work, we propose a diffusion-based model that generates realistic and long-form dance sequences conditioned on music.
We evaluate our model on multiple automated metrics (including our proposed PFC) and a large user study, and find that it achieves state-of-the-art results on the AIST++ dataset and generalizes well to in-the-wild music inputs.
Importantly, we demonstrate that our model admits powerful editing capabilities, allowing users to freely specify both temporal and joint-wise constraints.
The introduction of editable dance generation provides multiple promising avenues for future work.
Whereas our method is able to create arbitrarily long dance sequences via the chaining of locally consistent, shorter clips, it cannot generate choreographies with very-long-term dependencies. Future work may explore the use of non-uniform sampling patterns such as those implemented in Harvey et al. \cite{harvey2022flexible}, or a variation of the frame inbetweening scheme used in Ho et al. \cite{ho2022imagen}.
Editability also opens the door to the generation of more complex choreographies, including multi-person and scene-aware dance forms.
We are excited to see the future that this new direction of editable dance generation will enable.

{\small
\bibliographystyle{ieee_fullname}
\bibliography{main}
}

\ifarxiv \clearpage \appendix
\definecolor{codegreen}{rgb}{0,0.6,0}
\definecolor{codegray}{rgb}{0.5,0.5,0.5}
\definecolor{codepurple}{rgb}{0.58,0,0.82}
\definecolor{backcolour}{rgb}{0.95,0.95,0.92}

\lstdefinestyle{mystyle}{
    backgroundcolor=\color{backcolour},   
    commentstyle=\color{codegreen},
    keywordstyle=\color{magenta},
    numberstyle=\tiny\color{codegray},
    stringstyle=\color{codepurple},
    basicstyle=\ttfamily\footnotesize,
    breakatwhitespace=false,         
    breaklines=true,                 
    captionpos=b,                    
    keepspaces=true,                 
    numbers=left,                    
    numbersep=5pt,                  
    showspaces=false,                
    showstringspaces=false,
    showtabs=false,                  
    tabsize=2
}

\lstset{style=mystyle}

\label{sec:appendix}

\section{User Study}
\label{sec:appendix_user_study}

Quality responses were ensured by three separate mechanisms:
\begin{itemize}[itemsep=-1ex,partopsep=1ex,parsep=1ex]
    \item Prolific's automatic vetting, which ensures that participants whose survey responses are rejected too often are kicked from the platform.
    \item Restriction to participants from the United States, which is known to mitigate the frequency of fraudulent responses~\cite{kennedy2020shape}.
    \item Filtering out participants via identifying survey responses which incorrectly answer control questions.
    By this process, we eliminated 20 participants, resulting in a filtered total of 147 participants.
\end{itemize}

\begin{figure}
    \centering
    \includegraphics[width=\linewidth]{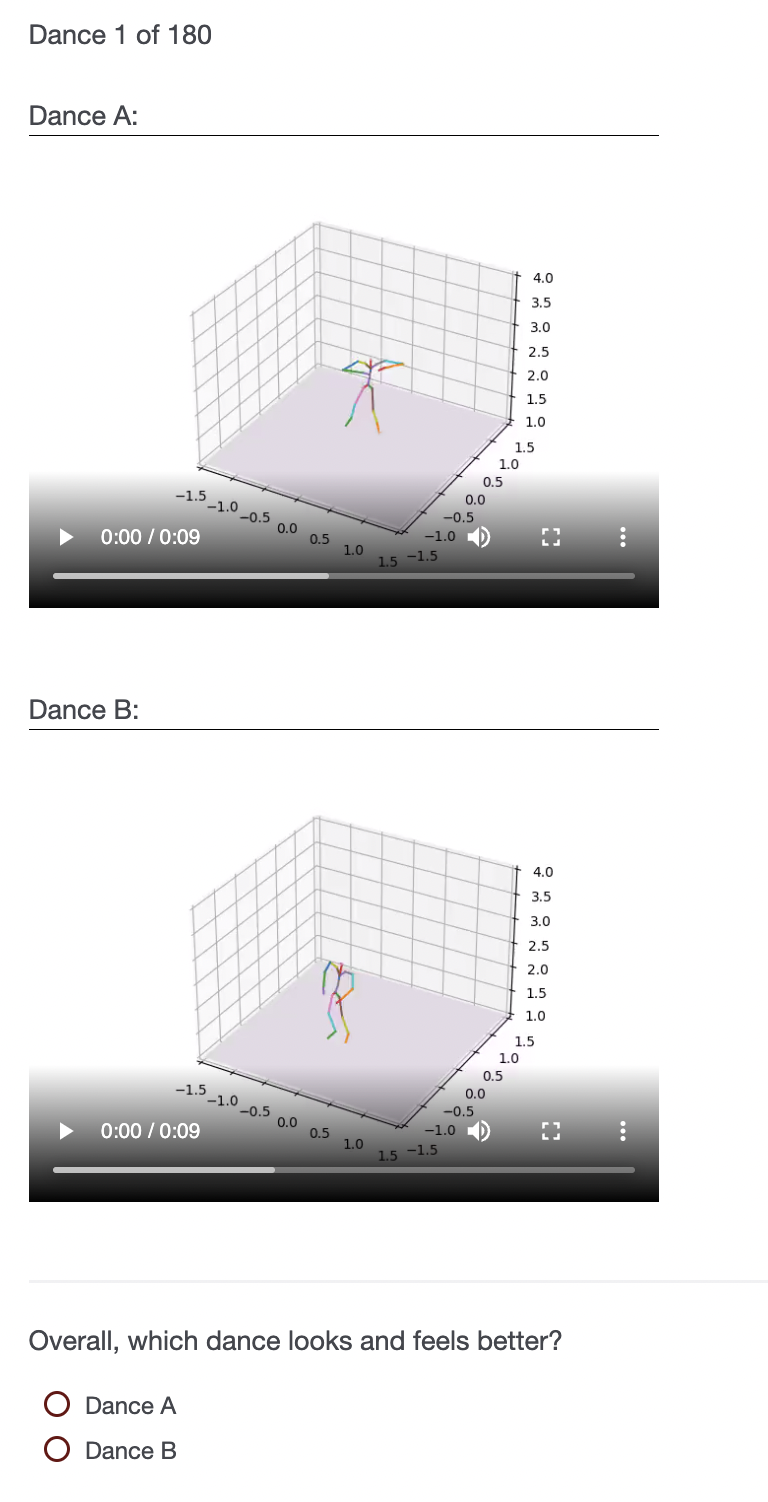}
    \caption{A screenshot of the user interface from our surveys.
    We use simple stick figures as the skeleton for our dance to (1) ensure fair comparisons with \textit{Bailando}, which is difficult to rig due to the fact that it operates in Cartesian space, and (2) computational efficiency.}
    \label{fig:6}
\end{figure}

For our overall evaluations, we also evaluated against the 10 checkpoints taken throughout model training in order to increase coverage across skill levels for our Elo calculations.

In our physical plausibility survey, we ask ``Which dance is more physically plausible?'', but otherwise the survey is exactly the same.

Over the course of 24 days, we collected responses from all of the raters and plot them below in a win table. We compute our Elo and win rate metrics from this table's numbers.
Elo was computed using the \href{https://github.com/ddm7018/Elo}{Elo package} in Python by randomly sorting all matchups and obtaining the average Elo over 1000 trials.
See \cref{tab:win_table} for the raw win numbers across all of the methods we tested.

For our confidence intervals, we treated each pair of methods independently, and then each win or fail as an independent Bernoulli trial, computing the 95\% confidence interval based on the number of wins and fails for that pair (Binomial distribution).

\begin{table*}[t]
\centering
\resizebox{\textwidth}{!}{
  \begin{tabular}{lcccccccccccccccccccccccc}
    \toprule
    \textbf{Method A / Method B} & \rotatebox[origin=c]{90}{Checkpoint 200} & \rotatebox[origin=c]{90}{Checkpoint 400} & \rotatebox[origin=c]{90}{Checkpoint 600} & \rotatebox[origin=c]{90}{Checkpoint 800} & \rotatebox[origin=c]{90}{Checkpoint 1000} & \rotatebox[origin=c]{90}{Checkpoint 1200} & \rotatebox[origin=c]{90}{Checkpoint 1400} & \rotatebox[origin=c]{90}{Checkpoint 1600} & \rotatebox[origin=c]{90}{Checkpoint 1800} & \rotatebox[origin=c]{90}{Checkpoint 2000} & \rotatebox[origin=c]{90}{Ground Truth} & \rotatebox[origin=c]{90}{FACT} & \rotatebox[origin=c]{90}{EDGE (J+C' ($w=2$))} & \rotatebox[origin=c]{90}{EDGE (J+C ($w=2$))} & \rotatebox[origin=c]{90}{EDGE (J ($w=2$))} & \rotatebox[origin=c]{90}{EDGE ($w=2$)} & \rotatebox[origin=c]{90}{\textit{Bailando}} & \rotatebox[origin=c]{90}{EDGE (J+C OOD ($w=2$))} & \rotatebox[origin=c]{90}{EDGE (C OOD ($w=2$))} & \rotatebox[origin=c]{90}{EDGE (J+C ($w=1$))} & \rotatebox[origin=c]{90}{EDGE (J ($w=1$))} & \rotatebox[origin=c]{90}{EDGE (J+C' ($w=1$))} & \rotatebox[origin=c]{90}{FACT (OOD)} & \rotatebox[origin=c]{90}{\textit{Bailando} (OOD)} \\
    \midrule
    Checkpoint 200 & 0 & 3 & 3 & 3 & 1 & 2 & 0 & 1 & 0 & 0 & 1 & 0 & 0 & 0 & 0 & 1 & 0 & 0 & 1 & 0 & 0 & 0 & 0 & 0 \\
    Checkpoint 400 & 37 & 0 & 4 & 0 & 1 & 0 & 0 & 2 & 1 & 1 & 0 & 0 & 0 & 0 & 0 & 0 & 0 & 0 & 1 & 0 & 0 & 0 & 1 & 2 \\
    Checkpoint 600 & 37 & 36 & 0 & 4 & 3 & 0 & 1 & 2 & 1 & 1 & 2 & 7 & 0 & 1 & 2 & 0 & 0 & 0 & 0 & 0 & 1 & 1 & 1 & 16 \\
    Checkpoint 800 & 37 & 40 & 36 & 0 & 9 & 5 & 5 & 0 & 4 & 0 & 1 & 10 & 1 & 4 & 1 & 2 & 1 & 0 & 1 & 0 & 2 & 1 & 7 & 16 \\
    Checkpoint 1000 & 39 & 39 & 37 & 31 & 0 & 11 & 1 & 7 & 9 & 4 & 4 & 19 & 6 & 5 & 6 & 2 & 16 & 3 & 1 & 0 & 1 & 0 & 15 & 17 \\
    Checkpoint 1200 & 38 & 40 & 40 & 35 & 29 & 0 & 17 & 13 & 13 & 8 & 9 & 21 & 6 & 11 & 11 & 14 & 22 & 7 & 4 & 0 & 7 & 1 & 27 & 24 \\
    Checkpoint 1400 & 40 & 40 & 39 & 35 & 39 & 23 & 0 & 14 & 11 & 11 & 20 & 27 & 5 & 43 & 14 & 18 & 33 & 21 & 16 & 0 & 7 & 5 & 29 & 34 \\
    Checkpoint 1600 & 39 & 38 & 38 & 40 & 33 & 27 & 26 & 0 & 19 & 15 & 18 & 25 & 5 & 34 & 21 & 20 & 20 & 13 & 8 & 0 & 5 & 5 & 26 & 33 \\
    Checkpoint 1800 & 40 & 39 & 39 & 36 & 31 & 27 & 29 & 21 & 0 & 29 & 22 & 28 & 8 & 35 & 26 & 30 & 30 & 4 & 21 & 0 & 4 & 5 & 30 & 30 \\
    Checkpoint 2000 & 40 & 39 & 39 & 40 & 36 & 32 & 29 & 25 & 11 & 0 & 12 & 23 & 7 & 49 & 26 & 18 & 29 & 5 & 19 & 0 & 8 & 8 & 31 & 28 \\
    Ground Truth & 27 & 28 & 26 & 27 & 24 & 19 & 8 & 10 & 6 & 16 & 0 & 66 & 0 & 24 & 0 & 0 & 0 & 0 & 0 & 404 & 0 & 0 & 0 & 0 \\
    FACT & 28 & 28 & 21 & 18 & 9 & 7 & 1 & 3 & 0 & 5 & 4 & 0 & 0 & 7 & 0 & 0 & 0 & 0 & 0 & 137 & 0 & 0 & 0 & 0 \\
    EDGE (J+C' ($w=2$)) & 14 & 14 & 14 & 13 & 8 & 8 & 9 & 9 & 6 & 7 & 0 & 0 & 0 & 0 & 58 & 0 & 0 & 0 & 0 & 0 & 71 & 62 & 0 & 0 \\
    EDGE (J+C ($w=2$)) & 48 & 48 & 47 & 44 & 43 & 85 & 53 & 62 & 61 & 47 & 46 & 63 & 0 & 0 & 49 & 53 & 82 & 42 & 65 & 383 & 0 & 0 & 167 & 154 \\
    EDGE (J ($w=2$)) & 32 & 32 & 30 & 31 & 26 & 39 & 36 & 29 & 24 & 24 & 0 & 0 & 82 & 41 & 0 & 45 & 87 & 0 & 0 & 446 & 83 & 80 & 0 & 0 \\
    EDGE ($w=2$) & 17 & 18 & 18 & 16 & 16 & 22 & 18 & 16 & 6 & 18 & 0 & 0 & 0 & 37 & 45 & 0 & 85 & 0 & 0 & 353 & 0 & 0 & 0 & 0 \\
    \textit{Bailando} & 18 & 18 & 18 & 17 & 2 & 14 & 3 & 16 & 6 & 7 & 0 & 0 & 0 & 8 & 3 & 5 & 0 & 0 & 0 & 193 & 0 & 0 & 0 & 0 \\
    EDGE (J+C OOD ($w=2$)) & 13 & 13 & 13 & 13 & 10 & 19 & 5 & 13 & 22 & 21 & 0 & 0 & 0 & 36 & 0 & 0 & 0 & 0 & 57 & 0 & 0 & 0 & 0 & 0 \\
    EDGE (C OOD ($w=2$)) & 12 & 12 & 13 & 12 & 12 & 22 & 10 & 18 & 5 & 7 & 0 & 0 & 0 & 13 & 0 & 0 & 0 & 21 & 0 & 0 & 0 & 0 & 0 & 0 \\
    EDGE (J+C ($w=1$)) & 0 & 0 & 0 & 0 & 0 & 0 & 0 & 0 & 0 & 0 & 256 & 523 & 0 & 277 & 214 & 307 & 467 & 0 & 0 & 0 & 0 & 0 & 0 & 0 \\
    EDGE (J ($w=1$)) & 14 & 14 & 13 & 12 & 13 & 7 & 7 & 9 & 10 & 6 & 0 & 0 & 69 & 0 & 57 & 0 & 0 & 0 & 0 & 0 & 0 & 75 & 0 & 0 \\
    EDGE (J+C' ($w=1$)) & 14 & 14 & 13 & 13 & 14 & 13 & 9 & 9 & 9 & 6 & 0 & 0 & 78 & 0 & 60 & 0 & 0 & 0 & 0 & 0 & 65 & 0 & 0 & 0 \\
    FACT (OOD) & 17 & 16 & 16 & 10 & 2 & 7 & 5 & 8 & 4 & 3 & 0 & 0 & 0 & 20 & 0 & 0 & 0 & 0 & 0 & 0 & 0 & 0 & 0 & 73 \\
    \textit{Bailando} (OOD) & 17 & 15 & 1 & 1 & 0 & 10 & 0 & 1 & 4 & 6 & 0 & 0 & 0 & 33 & 0 & 0 & 0 & 0 & 0 & 0 & 0 & 0 & 29 & 0 \\
    \bottomrule
\end{tabular}}
\caption{This table shows the total number of wins of Method A against Method B across our large-scale user study. For the method names, ``J'' means Jukebox features, ``C'' means CCL, ``$C'$'' means an early CCL prototype, $w$ means guidance weight, OOD means out-of-domain (in-the-wild), and each ``Checkpoint N'' method refers to the checkpoint taken from epoch N of training, as seen in \cref{fig:5}.}
\label{tab:win_table}
\end{table*}

\section{FACT and \textit{Bailando}}

We use the \href{https://github.com/google-research/mint/}{official implementation} for our FACT experiments.
We noted that the published FACT checkpoints were \href{https://github.com/google-research/mint/issues/42#issuecomment-1066211911}{not fully trained}, so we re-trained the model from scratch for $300k$ steps, following the original paper \cite{li2021ai}.
After inference, we downsample the model predictions to 30 FPS and perform forward kinematics.
We also use the \href{https://github.com/lisiyao21/Bailando}{official \textit{Bailando} implementation} and their published checkpoints for inference.
After inference, we downsample the predictions to 30 FPS and render the joint positions directly.

\section{Metrics and Evaluation}
\label{sec:appendix_metrics_and_eval}

For automatic evaluation (PFC, beat alignment, Dist$_k$, and Dist$_g$), we sampled 5-second clips from each model, using slices taken from the test music set sampled with a stride of 2.5 seconds as music input.
For qualitative evaluation via the Prolific study, we sampled 10-second clips from each model, using slices taken from the test music set sampled with a stride of 5 seconds as music input.

To extract features for FID and diversity results, we directly use the official
\href{https://github.com/google-research/mint}{FACT GitHub repository}, which borrows code from \href{https://github.com/facebookresearch/fairmotion}{fairmotion}.

\section{PFC}
\label{sec:appendix_pfc}

We explain two implementation details of PFC:
\begin{enumerate}
    \item The PFC equation (\cref{eq:10}) depends on the acceleration of the center of mass; since masses are not explicitly annotated in the AIST++ dataset, we use the acceleration of the root joint as a practical approximation.
    \item The PFC equation (\cref{eq:10}) normalizes only the COM acceleration, and not the foot velocities. Under the assumptions of static contact, a body with at least one static foot is able to generate arbitrary (within physical reason) amounts of COM acceleration. Under these assumptions, two sequences where the root acceleration differs but the foot velocity is the same are \textit{equally plausible}. In contrast, two sequences with the same root acceleration but different foot velocities are not equally plausible.
\end{enumerate}

We perform an analysis in which we take the same checkpoints as from \cref{par:fidk} and evaluate their PFC scores.

As can be seen from \cref{fig:7}, PFC scores tend to improve over the course of training, providing further evidence that this metric measures physical plausibility.

\begin{figure}
    \centering
    \includegraphics[width=\linewidth]{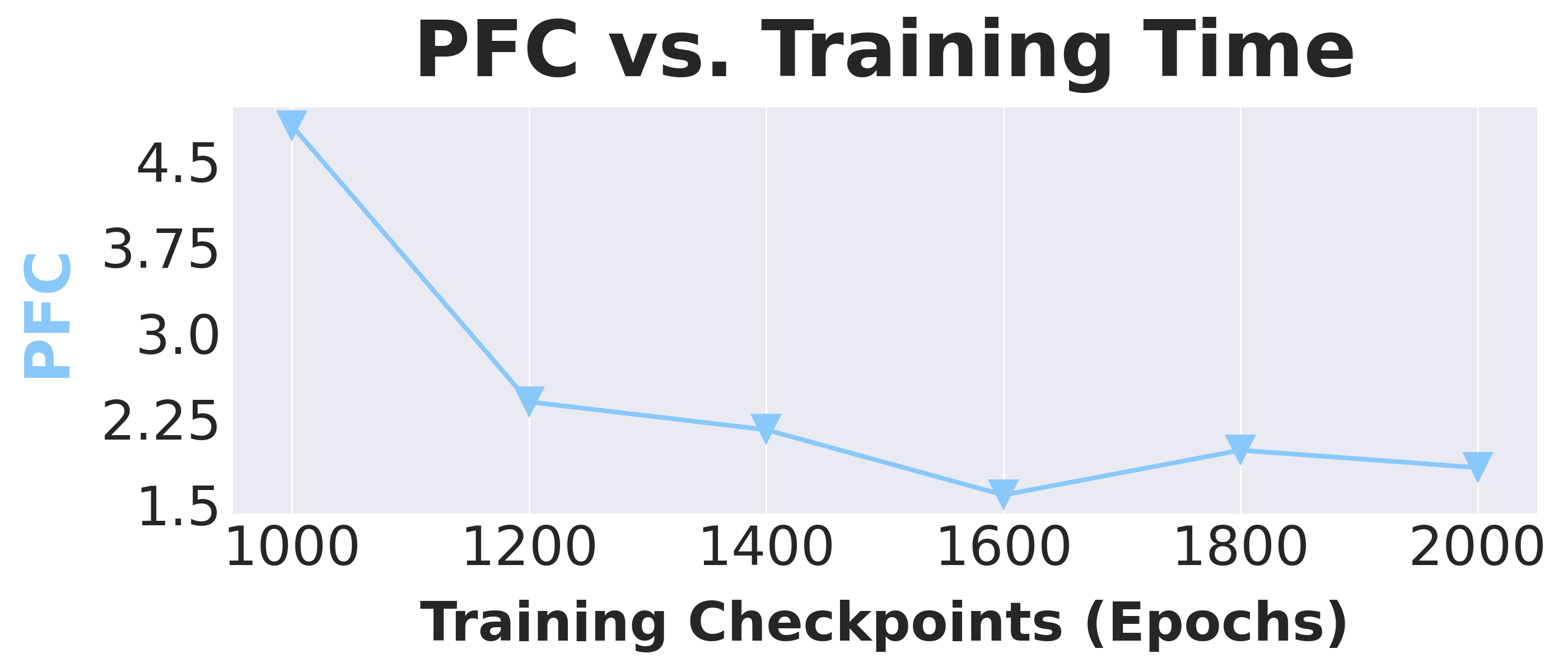}
    \caption{PFC decreases throughout training.}
    \label{fig:7}
\end{figure}

\section{In-the-wild Music}
\label{sec:appendix_ood_music}

We use the below songs for our in-the-wild music demos:
\begin{itemize}
    \item \href{https://www.youtube.com/watch?v=S05hfsRbw3M}{Doja Cat - Woman}
    \item \href{https://www.youtube.com/watch?v=kJQP7kiw5Fk}{Luis Fonsi - Despacito ft. Daddy Yankee}
    \item \href{https://www.youtube.com/watch?v=MjCZfZfucEc}{ITZY - LOCO}
    \item \href{https://www.youtube.com/watch?v=x5c2iRHlAHA}{Saweetie - My Type}
    \item \href{https://www.youtube.com/watch?v=Xv9023cub0Y}{Rihanna - Only Girl (In The World)})
\end{itemize}

\section{Hyperparameters}

\begin{tabular}{c|c} 
 \hline
 Hyperparameter & Value \\
 \hline
 Optimizer & Adan~\cite{xie2022adan} \\ 

 Learning Rate & 4e-4 \\

 Diffusion Steps & 1000 \\

 $\beta$ schedule & cosine \\

 Motion Duration & 5 seconds\\

 Motion FPS & 30\\

 Motion Dimension & 151\\

 Classifier-Free Dropout & 0.25\\

 Num Heads & 8\\

 Num Layers & 8\\

 Transformer Dim & 512\\

 MLP Dim & 1024\\

 Dropout & 0.1\\

 EMA Steps & 1\\

 EMA Decay & 0.9999\\
 \hline
\end{tabular}

\section{Guidance Weight at Inference Time}
In our experiments, we find that dropping out the guidance at early denoising steps (i.e. set $w=0$ from step 1000 to step 800) further helps to increase diversity.
We perform this dropout for the version of our model sampled at $w=1$, dropping out 40\% of the steps.

\section{Memory-efficient Jukebox Implementation}
\label{sec:appendix_juke_mem}

Jukebox extraction implementations from previous work \cite{castellon2021codified,donahue2021sheet} are limited in memory efficiency (cannot load in the full model on a GPU with 16GB VRAM) and speed on short clips (performs inference as if the clip has full sequence length).

We improve upon these implementations and develop a new implementation that (1) improves memory efficiency two-fold and (2) speeds up extraction time 4x for 5-second clips.

\begin{itemize}
    \item For memory efficiency, we use \href{https://huggingface.co/docs/accelerate/v0.11.0/en/big_modeling}{HuggingFace Accelerate} to initialize the model on the ``meta'' device followed by loading in the checkpoint with CUDA, meaning that only half of the memory is used (using the traditional loading mechanism, PyTorch does not de-allocate the initial random weights allocated before the checkpoint is loaded in).
    \item For speed, we adapt the codebase to accept and perform inference on shorter clips.
    Our new implementation takes only $\sim\hspace{-3pt}5$ seconds for a 5-second clip on a Tesla T4 GPU.
\end{itemize}

These improvements make extracting representations from Jukebox more accessible to researchers and practitioners.

Additionally, to downsample to 30 FPS we use \textit{librosa}'s \href{https://librosa.org/doc/main/generated/librosa.resample.html}{resampling method} with ``fft'' as the resampling type to avoid unnecessary slowness.

We plan to release this new implementation together with our code release.

\section{Long-Form Generation}

Shown below is the pseudocode for long-form sampling.

\begin{lstlisting}[language=Python]
def long_generate():
    z = randn((batch, sequence, dim))
    half = z.shape[1] // 2
    for t in range(0, num_timesteps)[::-1]:
        # sample z from step t to step t-1
        z = p_sample(z, cond, t)
        # enforce constraint
        if t > 0:
            z[1:,:half] = z[:-1,half:]
    x = z
    return x
\end{lstlisting}

\section{Rigging}
In order to render generated dances in three dimensions for our \website, we export joint angle sequences to the FBX file format to be imported into Blender, which provides the final rendering. The character avatar used in all our renders except for those of dances from \textit{Bailando} is ``Y-Bot'', downloaded from Mixamo. Dances from \textit{Bailando} are rendered using ball-and-rod stick figures to ensure fair comparison, since dances are generated in Cartesian joint position space. Although inverse kinematic (IK) solutions are available, we sought to avoid the potential introduction of extraneous artifacts caused by poorly tuned IK, and instead rendered the joint positions directly. We perform no post-processing on model outputs.
\section{Extra Figures}
\begin{figure*}
    \centering
    \includegraphics[width=0.9\linewidth]{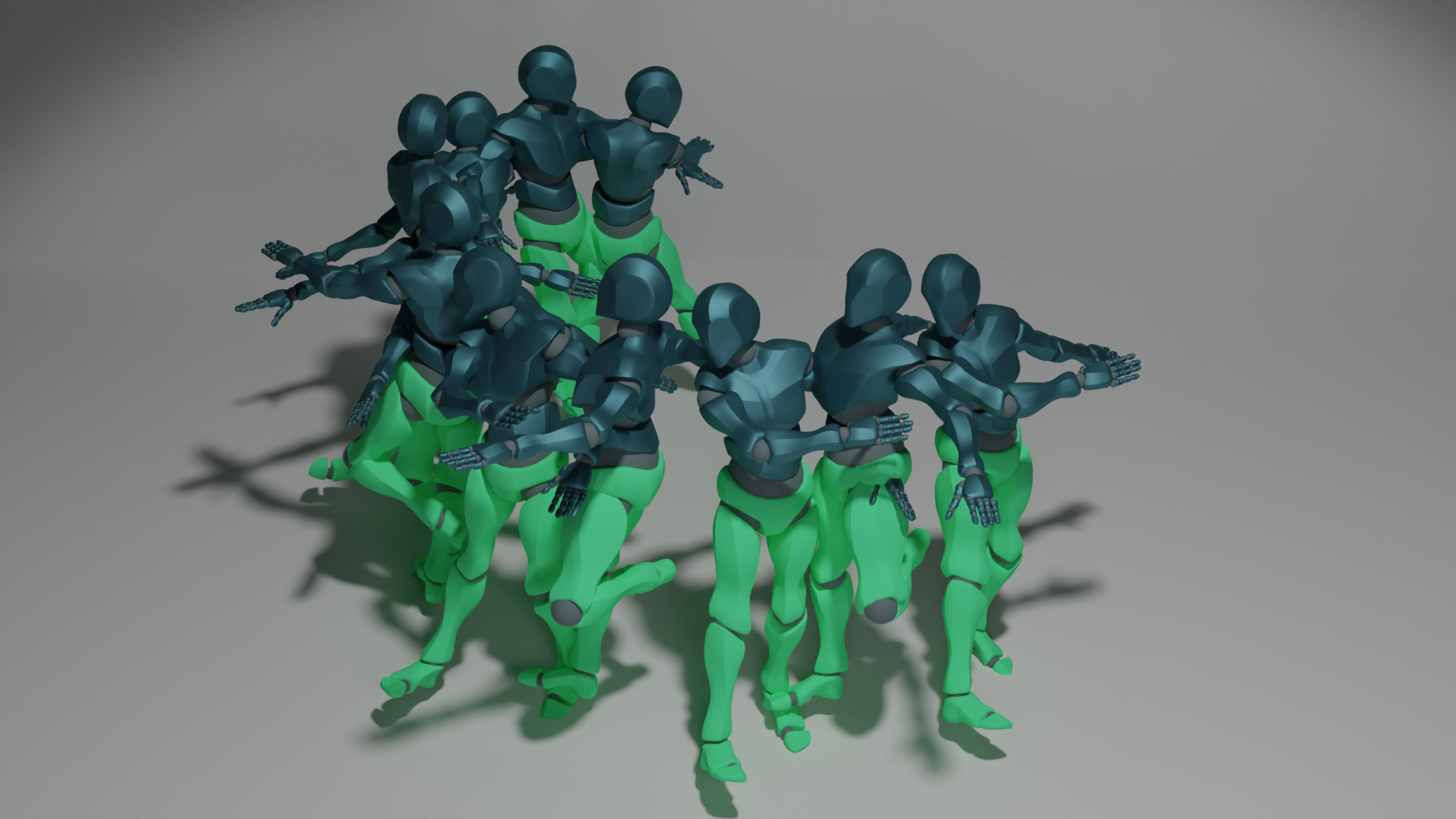}
    \caption{\textbf{Joint-wise Conditioning:} EDGE can generate upper body motion given a lower body motion constraint.}
    \label{fig:8}
\end{figure*}
\begin{figure*}
    \centering
    \includegraphics[width=0.9\linewidth]{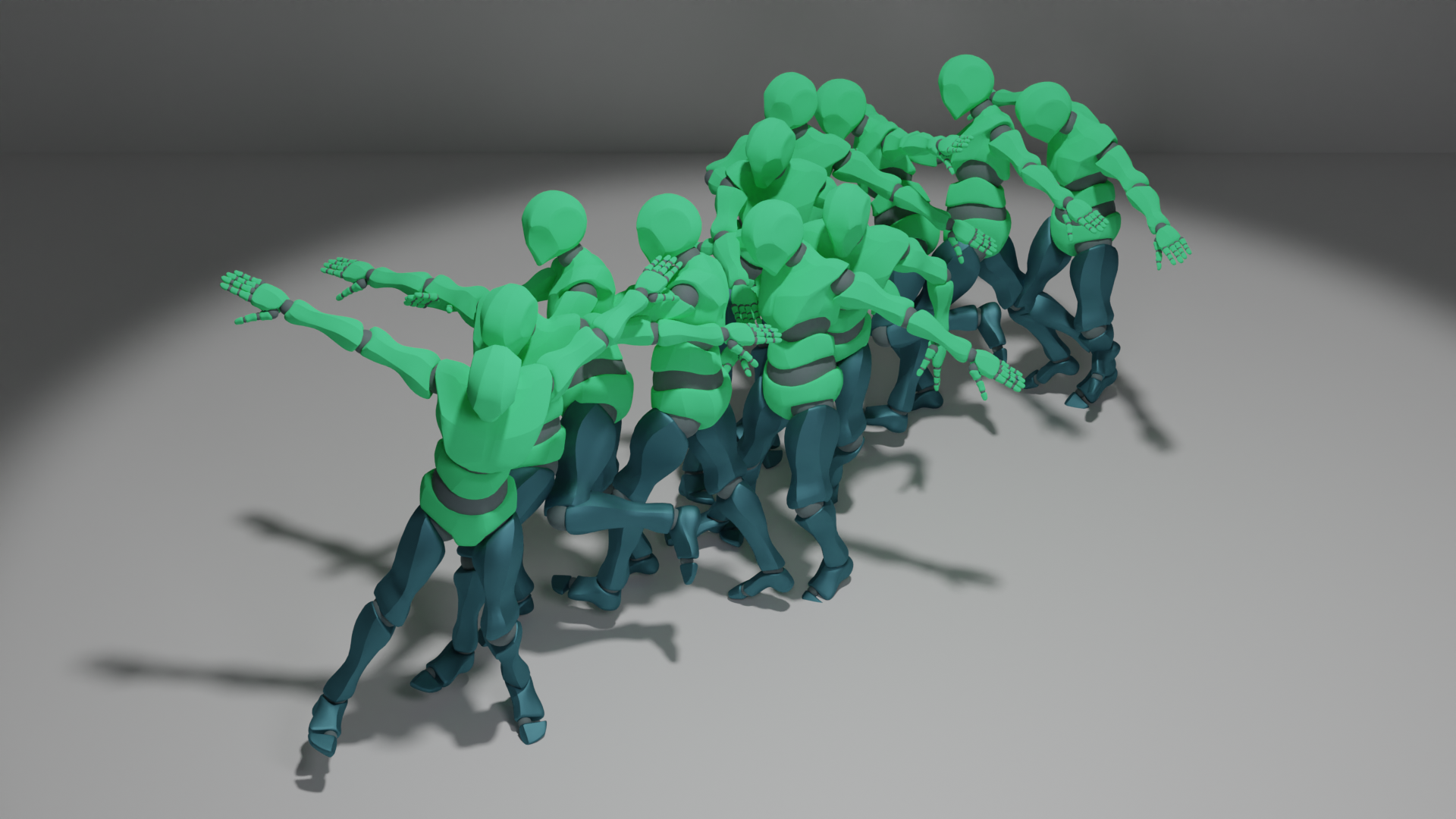}
    \caption{\textbf{Joint-wise Conditioning:} EDGE can generate lower body motion given an upper body motion constraint.}
    \label{fig:9}
\end{figure*}
\begin{figure*}
    \centering
    \includegraphics[width=0.9\linewidth]{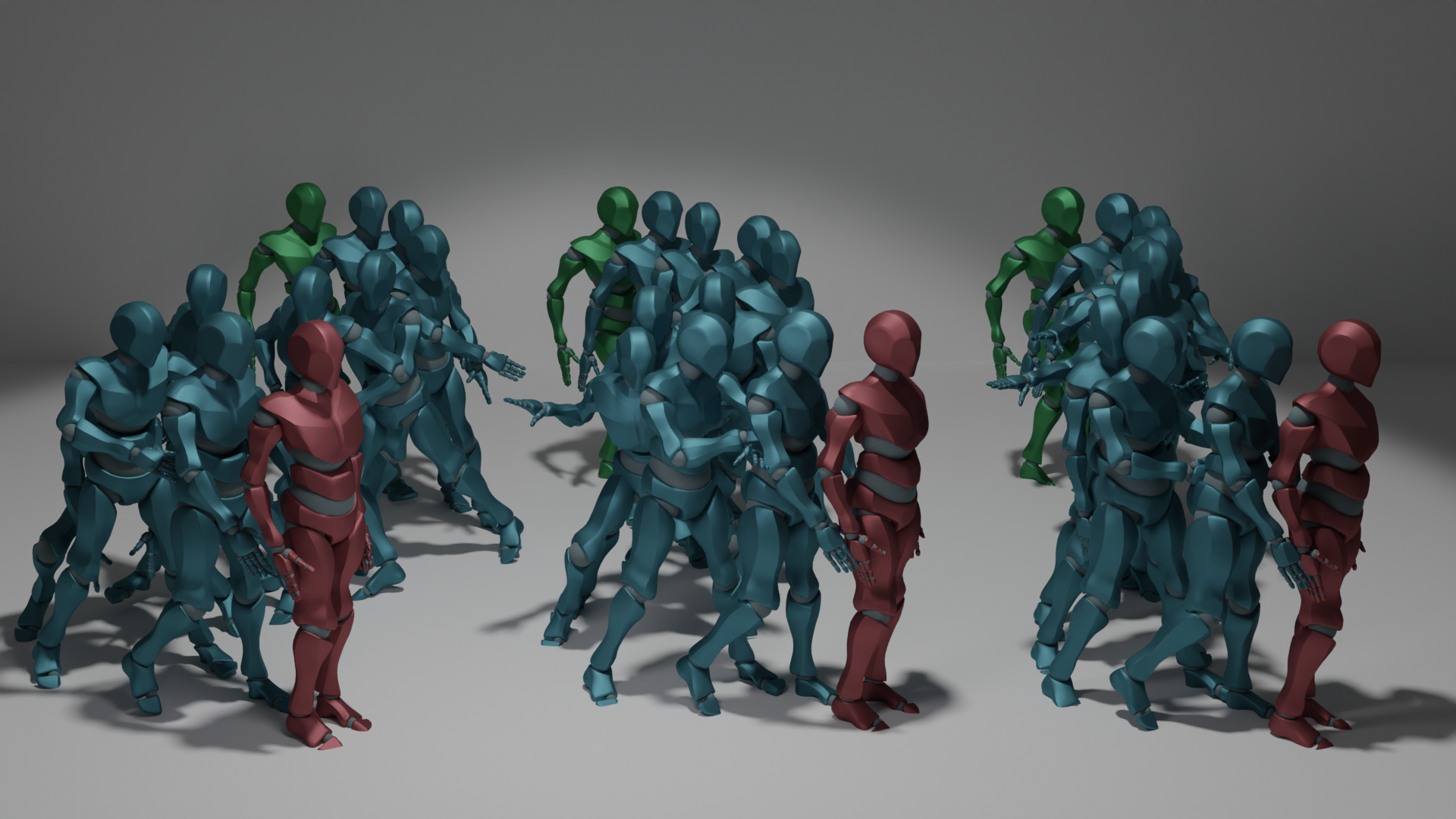}
    \caption{\textbf{Temporal Conditioning: In-Betweening} Given start and end poses, EDGE can generate diverse in-between dance sequences.}
    \label{fig:10}
\end{figure*}
\begin{figure*}
    \centering
    \includegraphics[width=0.9\linewidth]{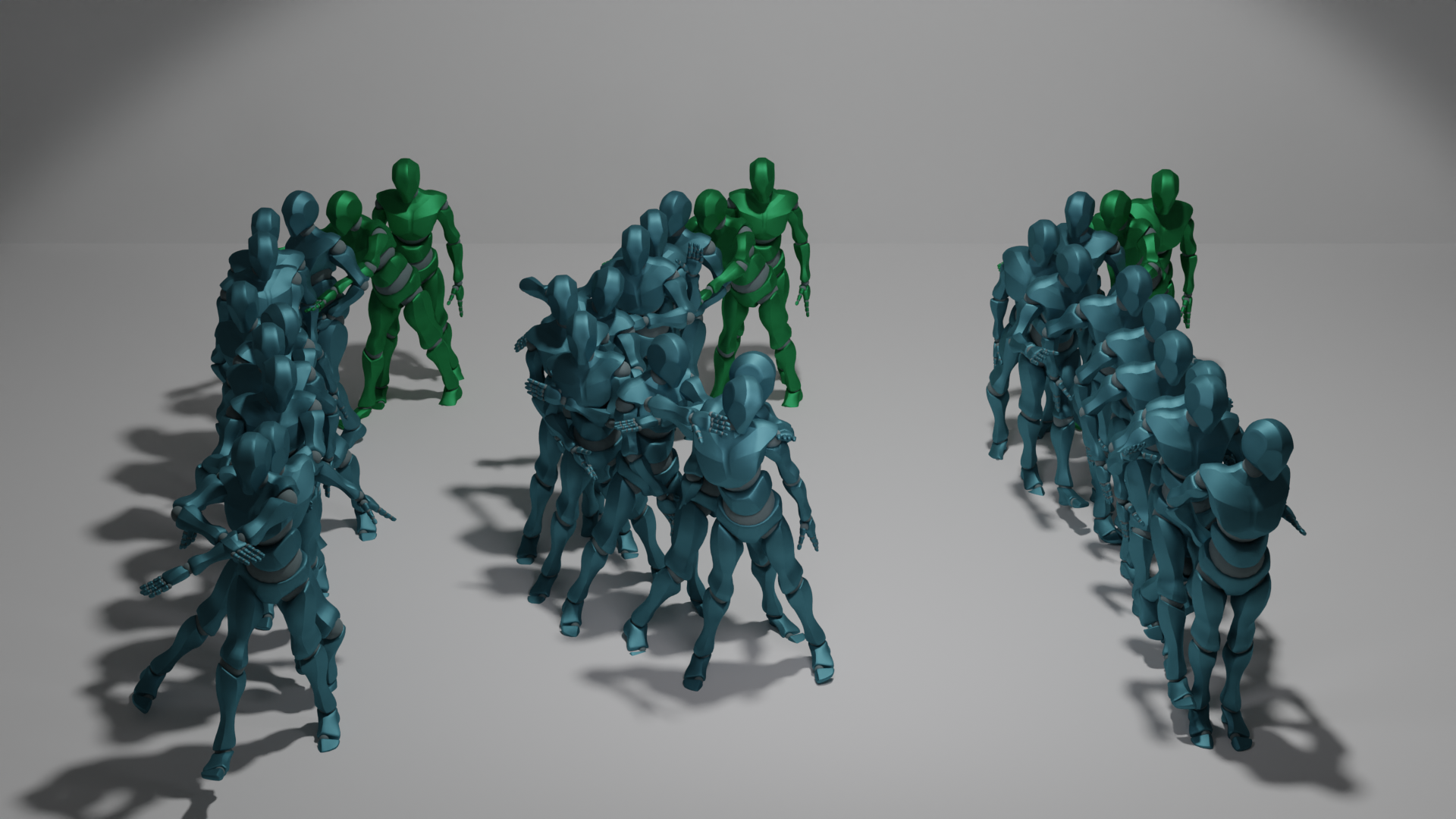}
    \caption{\textbf{Temporal Conditioning: Continuation} Given a seed motion, EDGE can continue the dance  in response to arbitrary music conditioning.}
    \label{fig:11}
\end{figure*} \fi

\end{document}